\documentclass[doublecol,figures]{epl2}
\usepackage{graphicx}
\usepackage[english]{babel}
\newcommand{\be}{\begin{equation}}
\newcommand{\ee}{\end{equation}}

\begin{document}

\hyphenation{de-ca-de}
\hyphenation{gra-nu-lar}
\hyphenation{ba-lan-ce}
\hyphenation{con-ti-nui-ty}
\hyphenation{con-si-de-ring}
\hyphenation{ve-lo-ci-ty}
\hyphenation{de-ve-lo-ped}
\title{Shear bands in granular flow through a mixing length model}
\author{Riccardo Artoni\thanks{E-mail: \email{riccardo.artoni@unipd.it}} \and Andrea Santomaso \and Paolo Canu
%
}                     
\shortauthor{R. Artoni \etal}
%
%
\institute{Dipartimento di Principi e Impianti di Ingegneria Chimica ``I. Sorgato''\\Universit\`{a} di Padova. Via Marzolo 9, 35100 Padova, Italy.}

\abstract{We discuss the advantages and results of using a mixing-length, compressible model to account for shear banding behaviour in granular flow. We formulate a general approach based on two function of the solid fraction to be determined. Studying the vertical chute flow, we show that shear band thickness is always independent from flowrate in the quasistatic limit, for Coulomb wall boundary conditions. The effect of bin width is addressed using the functions developed by Pouliquen and coworkers, predicting a linear dependence of shear band thickness by channel width, while literature reports contrasting data. We also discuss the influence of wall roughness on shear bands. Through a Coulomb wall friction criterion we show that our model correctly predicts the effect of increasing wall roughness on the thickness of shear bands. Then a simple mixing-length approach to steady granular flows can be useful and representative of a number of original features of granular flow.}
\maketitle
%
\section{Introduction}
\label{intro}
Granular materials are ubiquitous in everyday's life, but a satisfying comprehension of their complex dynamics has not yet been achieved, despite the efforts of condensed matter science\cite{jaeger}. From the latin poet Lucretius to nowadays, we have experienced a perspective reversal; while, in his De Rerum Natura, he used seeds as an example to explain the ``flowability'' of liquids, today we try and use hydrodynamic analogies (\textit{e.g.} \cite{joss04},\cite{savage},\cite{bocquet},\cite{losert}) to explain the flow of granular materials. However, towards those poppy seeds our various theories remain almost a divinatory exercise.\\
Many of the theoretical approaches appeared in last decade literature neglect the compressibility of granular materials, assuming it as an incompressible fluid with $\rho\approx const$. From a phenomenological point of view, we believe that dilatancy is a requirement for shearing a granular material, in other words, the material has to dilate in order to shear. Accordingly, we expect that neglect of dilatancy loses an important part of the granular flow physics.\\
In this work we formulate a simple model explicitly involving the solid fraction influence on the flow properties of granular materials. The model aim at being a generalization of the one by Pouliquen et al.\cite{jop}\cite{poulipow05} based on the dimensionless parameter $I$, considered as the ratio between shearing time and rearranging time due to pressure. In Pouliquen's formulation, the solid fraction is derived from $I$ as being linearly depending on it. Here we reverse the formulation assuming the solid fraction $\phi$ to be the critical variable, instead of $I$, to reestablish the physical relevance of the dilation of the medium to determine the flow features. Giving the model an appropriate account of the solid fraction can become important for those geometries (like silos) in which $\phi$ varies significantly (more than $10 \%$) all over the flow section. In silos flow can be seen to originate from fluidization due to the injection of voids from the exit hole, where solid fraction is quite different from its value in the core.\\
In this perspective, we illustrate the $\phi$-based model derivation and its application to the vertical chute arrangement, to verify the constitutive relations proposed.

\section{The model}
The model outlined here is formulated for 2D, steady granular flows. The relevant equations are momentum balance with its two components, and the equation of continuity.\\ 
As a fundamental assumption, we consider the flow structure to be solvable with the steady, compressible Navier-Stokes (N-S) equations, with a solid fraction, pressure and shear-dependent viscosity. n addition, we need a constitutive equation for $\eta$, and we saturate the degree of freedom introduced by $\phi$ with a sort of Equation of State that involve pressure.\\
Accordingly, equations are:\\
\be
	\nabla \cdot \rho\vec{u}=0
	\label{mod1}
\ee
\be
 \nabla \cdot (\rho u \vec{u})=-\frac{\partial p}{\partial x}+2\frac{\partial\; }{\partial x} \eta \left( \frac{\partial u }{\partial x}\right)+\frac{\partial\; }{\partial y} \eta \left( \frac{\partial u }{\partial y}+\frac{\partial v }{\partial x}\right)+\rho g_x
	\label{mod2}
\ee
\be
\nabla \cdot (\rho v \vec{u})=-\frac{\partial p}{\partial y}+2\frac{\partial\; }{\partial y} \eta \left( \frac{\partial v }{\partial y}\right)+\frac{\partial\; }{\partial x} \eta \left( \frac{\partial u }{\partial y}+\frac{\partial v }{\partial x}\right)+\rho g_y
\label{mod3}
\ee
where $\rho$ is the local density of the compressible pseudo-homogeneous medium, which indeed is a two phase mixture:
\be
\rho=\rho_p \phi+\rho_{int} (1-\phi)
\ee
being $\rho_p$ and $\rho_{int}$material and interstitial fluid densities, respectively. If the interstitial fluid is air, given that $\rho_{air} < < \rho_p$, we can approximate
\be
\rho=\rho_p \phi
\ee

\subsection{Constitutive relations}
We assume a rheological law based on dimensional analysis, like Prandtl's approach to turbulent flows, as discussed by Ertas and Halsey\cite{ertas}. Because of the eminent precursor, we will call it ``mixing length approach''.\\
In this perspective the apparent viscosity of the medium is locally formulated as:
\be
\eta=\rho_p\;L^2\;\left|\dot{\gamma}\right|
\label{eta}
\ee
where the unique timescale is $\left|\dot{\gamma}\right|^{-1}$; $L$ is a characteristic length, that has to be function of $d$ and $\phi$ only, with a generic relation of the form: 
\be
L^2=d^2\;f(\phi)		\label{L2}
\ee
The function $f(\phi)$ is not known so far, but some features of it may be prescribed: it should diverge when $\phi\rightarrow\phi_{max}$, to limit the material flow (that becomes 'jammed'), i.e. $\eta\rightarrow\infty$. To achieve this limit $\eta$ should diverge faster than $\left|\dot{\gamma}\right|^{-1}$, as it can be easily seen from Eq. \ref{eta}. Interestingly, for $f(\phi)=1$ Eq. \ref{L2} reduce to Bagnold scaling for shear stress (valid for rapid granular flows), providing a further requirement that $f(\phi\rightarrow 0)=1$. However, the present work addresses dense flow of granular material, and we are not interested, at the moment, in the liquid-gas like transition.\\
We also need a relation between pressure and solid fraction, which is similar to an Equation of State (EoS). Assuming shear rate plays the role of temperature in a gas, and $\phi$ acts through a geometrical (excluded volume) function $h(\phi)$ to be specified, we can use dimensional analysis the obtain the following EoS:
\be
p=\rho_p\;h(\phi)\;(\left|\dot{\gamma}\right|\;d)^2
\label{EoS}
\ee
To keep pressure finite when shear rate vanishes, $h(\phi)$ has to diverge when $\phi\rightarrow\phi_{max}$ (although value and physical meaning of $\phi_{max}$ is still a matter of debate \cite{joss06}).\\
Dimensional analysis is broadly used in granular flow modeling attempts, starting from Bagnold's works\cite{bagnold}. The formulation of Josserand et al. \cite{joss04}\cite{joss06} uses dimensional analysis with Coulomb friction to develop a constitutive relation for shear stress that is composed by a rate dependent part and a rate independent one, and where the isotropic part of the stress tensor is related to solid fraction by means of entropic considerations. We express normal and shear stresses according to Pouliquen \cite{jop}\cite{poulipow05}, with the difference that we use explicitly solid fraction as the key variable, instead of dimensionless shear rate. Note that these laws, are very similar to those developed from hydrodynamic analogies \cite{bocquet}\cite{losert}, where granular temperature is used to represent the local mobility of the medium. We prefer our simple closure, based on $\phi$ and an EoS for it, also because granular temperature is a variable which is difficult to measure and then correctly validate.\\
Rearranging  Eq. \ref{eta}-\ref{L2}, we obtain:
\be
\eta=\frac{p}{\left|\dot{\gamma}\right|}\frac{f(\phi)}{h(\phi)}=\frac{p}{\left|\dot{\gamma}\right|}G(\phi)
\label{eta2}
\ee 
For sake of simplicity, we replaced the ratio $f/h$ with $G(\phi)$ and introduce:

\be
F(\phi)=[h(\phi)]^{-1/2}
\ee
as a simple replacement, provided $h$ always appears in this form in the following developments of NS eqs.
Note that $G$ must vanish if $\eta \dot{\gamma} = \tau \rightarrow 0$.

\section{Applying the model to the vertical chute}
We choose the vertical chute configuration as a standard benchmark for model evaluation. Original flow structures, principally related to the width of shear zones, can  be found in the chute flow, like in Couette cells. Reference for these configurations is the well known paper by GDR MiDi\cite{gdr}. Broadly speaking, the material flows in a plug-like fashion in the central part of the chute, while it is sheared near the wall. The extent of shear bands apparently approaches a typical dimension, of order 10-15 particle diameters. Predicting shear bands' thickness is a benchmark for all models applied to the chute and Couette flow\cite{poulipow05}.

\subsection{Vertical chute equations}
\begin{figure}[!t]
\centering{
	\resizebox{0.4\columnwidth}{!}{%
	\includegraphics{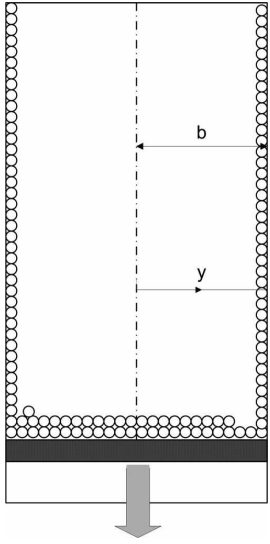}
	}}
	\caption{Vertical chute scheme.}
	\label{ciute}
\end{figure}

A scheme of the chute is given in Fig. \ref{ciute}.
For the steady vertical chute 2D flow, N-S equations simplify thanks to:
\be
	v=0\;\;\; \frac{\partial\;}{\partial x}=\frac{\partial\;}{\partial t}=0 
	\label{simp}
\ee

leading to:
\be
\left\{
	\begin{array}{ll}
	&\frac{\partial\; }{\partial y} \eta \left( \frac{\partial\; u }{\partial y}\right)+\rho_p \phi g=0\\
&\frac{\partial\; p}{\partial y} =0
		\label{chute2}
		\end{array}
		\right.
		\ee
We also included Janssen effect, assuming pressure does not vary in the vertical direction. Eq. \ref{chute2} additionally states that pressure will not vary horizontally either. The other equation can be integrated. With $y$ originating from the center of symmetry and directed to the wall, as shown in Fig. \ref{ciute}, we have:
\be
	\left|\dot{\gamma}\right|=-\left( \frac{\partial u }{\partial y}\right)
	\label{gp}
\ee
to be replaced in the $x$-momentum balance in combination with Eq. \ref{eta2}, to give:

\be
	\frac{\partial\; }{\partial y} \eta \left( \frac{\partial\; u }{\partial y}\right)+\rho_p \phi g=-p\frac{\partial  }{\partial y}\left[G(\phi)\right]+\rho_p \phi g=0
\ee
or

\be
	G(\phi) = \frac{\rho_p g}{p}\int^{y}_{0}\phi dy
	\label{chute3}
\ee
$G(y=0)=0$ follows from Eq. \ref{eta2} and symmetry, which requires the shear stress at the centerline to vanish.\\
From Eq. \ref{chute3} we expect to identify $\phi(y)$ provided $p$ and an invertible form of $G$ are given. At the same time, $u(y)$ can be obtained from 
the EoS:

\be
\left|\dot{\gamma}\right|^2=\frac{p}{\rho_p\;h(\phi)d^2}\Rightarrow 
 \frac{\partial u }{\partial y}=-\frac{\sqrt{p/\rho_p}}{d}F(\phi)
\ee
or
\be
u(y)=u_{slip}+\frac{\sqrt{p/\rho_p}}{d}\int_y^b F(\phi)\;dy
\label{chute4}
\ee
where $b$ is the half-width of the channel. 
So far, the unknown functions $\phi(y)$ and $u(y)$ can be formally obtained as: 
\be	\left\{
	\begin{array}{lcl}
	G(\phi)&=&\frac{\rho_p g}{p}\int^{y}_{0}\phi dy\\
	u(y) &=&u_{slip}+\frac{\sqrt{p/\rho_p}}{d}\int_y^bF(\phi)\;dy\\
	\end{array}
	\right.
	\label{resume}
\ee
but in practice $F(\phi)$ and $G(\phi)$ must be specified, and also the pressure calculated.

The continuity equation can be used in its integral form, correlating local profiles to the total mass flowrate, $\dot{M}$. Experiments are easily performed with constant flowrate, either controlled by a simple hole in the bottom of the silo, or using a moving plate with fixed velocity. Accordingly:

\be
2 \;\rho_p\int_{0}^{b}\phi u \;dy =\dot{M} =\mbox{const}
\label{tomare}
\ee
Developing, we can formulate an expression for the slip velocity by using Eq. \ref{chute4}:

\be
\int_{0}^{b} \phi \left(u_{slip}+\frac{\sqrt{p/\rho_p}}{d} \phi \int_y^bF(\phi) dy' \right) dy =\frac{\dot{M}}{2 \;\rho_p}
\ee

or:
\begin{eqnarray}
u_{slip}=\frac{1}{\int_{0}^{b} \phi dy}\left[\frac{\dot{M}}{2 \rho_p}-\frac{\sqrt{p/\rho_p}}{d}\int_{0}^{b} \phi \left(\int_y^bF(\phi)  dy'\right)  dy\right]
\label{tosorela}
\end{eqnarray}

\subsection{On boundary conditions}
One of the most critical issue in granular flow simulation is identification and application of boundary conditions. For the solid fraction, we will assume that in the central zone of the chute material reaches random close packing, or some other critical value of $\phi$ that leads pressure to diverge\cite{joss06}.\\
One constraint on velocity can be formulated as an integral condition, by fixing flowrate as done above.
In addition, we must speculate on the interaction between the granular assembly as a continuum and the walls. The simplest view used in literature is assuming a layer of particles glued at the walls, for which we can use a no-slip boundary condition. This assumption is attracting for its simplicity but we believe it requires caution in its application. We are skeptic that the continuum averaged interaction between nearest particles is the same in the bulk and in the layer of particles facing the glued ones. In this perspective, experimental investigation and critical theoretical speculations have to be done. A viable alternative to no-slip assumption is the Coulomb criterion at the wall:

\be
\tau_w= \sigma_w	tan \delta				\label{culomb}
\ee
where $\delta$ is a characteristic wall friction angle. In case of particle artificially fixed at the wall, this means assuming them as a wall, with a specific roughness measurable by its own $\delta$. 
Combining Coulomb's law with Eq. \ref{eta2} we obtain a condition on the solid fraction:

\be
\tau_w=\eta_w\left|\dot{\gamma}\right|_w=\frac{p}{\left|\dot{\gamma}\right|_w}G(\phi_w)\left|\dot{\gamma}\right|_w = p\;G(\phi_w)
\ee

which leads to
\be
G(\phi_w)=tan \delta
\label{ccphi}
\ee
given that $\sigma_w=p$. Very important, with Coulomb's criterion the slip velocity is not zero, and has to be determined from Eq. \ref{resume} using flowrate. 
In addition Eq. \ref{ccphi} allows to calculate the pressure (normal stress), provided that:

\be
G_w = G(y=b)=\frac{\rho_p g}{p}\int^{b}_{0} \phi \;dy
\ee
which gives:

\be
p=\frac{\rho_p g \int^{b}_{0}\phi\; dy}{tan \delta} = \frac{\rho_p \phi_{ave}\; g\; b }{tan \delta}
\label{presb}
\ee
where the average solid fraction, defined by $\int^{b}_{0}\phi dy = b \phi_{ave}$, has been introduced.

Combining Eqs. \ref{presb}, \ref{tosorela}, and \ref{resume}, the velocity profile can be explicitly written as:

\begin{eqnarray}\nonumber
u(y)&=&\frac{1}{\phi_{ave}b}\left[ -\frac{\sqrt{\phi_{ave}g b}}{d\sqrt{tan \delta}}\int_{0}^{b} \phi dy \left(\int_y^bF(\phi)dy \right) dy \right.\\
&& \left. + \frac{\dot{M}}{2 \rho_p} \right]+\frac{\sqrt{\phi_{ave}g b}}{d \sqrt{tan \delta}}\int_y^bF(\phi)dy
\end{eqnarray}
The result of Eq. \ref{presb} together with Eq. \ref{chute3}
leads to: 

\be
G(\phi)=\frac{tan \delta}{b\;\phi_{ave}}\int_{0}^{y}\phi\;dy 
\label{gtrue}
\ee
stating that in the approximation of $\phi\approx \phi_{ave}$

\be
G(\phi) \approx \frac{y}{b}\; tan \delta 
\label{gapprox}
\ee
or $G$ is a linear function of $y$, which provides a consistence criterion for the identification of the unknown function $G$.

Interestingly, our model, based also on Coulomb wall criterion, predicts the invariance of the velocity profiles with flowrate in the quasistatic limit (where $\phi\approx\phi_{ave}$ is valid). In other words, the scaled velocity profile:
 
\be
\tilde{u}=\frac{u(y)-u_{slip}}{u_{max}-u_{slip}}
=\frac{\int_{y}^{b}F(\phi(y))dy}{\int_0^{b}F(\phi(y))dy}
\label{shearbandw}
\ee
does not depend on flowrate, that influences only the slip tangential velocity. In this limit, the solid fraction profile also does not depend on flowrate. 
We underline that the result is independent of the particular formulation of the functions $F$ and $G$. Also with Pouliquen's formulation for the effective friction coefficient (which can be seen as a particular choice for $F$ and $G$), but with a Coulomb slip criterion, the independency of shear bands from flowrate is obtained. This is indeed a result supporting the mixing length approach. Profiles are flowrate-independent due to the nature of granular matter, which develops internal stresses to sustain itself; in this perspective, taking into account stresses in the formulation of boundary conditions is necessary.
However, far from the quasistatic limit, or conditions where $\phi\approx\phi_{ave}$ is acceptable, flowrate can significantly affect $\phi(y)$ and $u(y)$ because of the close coupling of the two equations.

\subsection{Deriving expressions for $F(\phi)$ and $G(\phi)$}
Following the work of Pouliquen and coworkers, the functions would take the form:
\be	\left\{
	\begin{array}{rcl}
	G(\phi)&=&\mu_s+\frac{\mu_2-\mu_s}{I_0/\tilde{\phi}+1}\\
	F(\phi)&=&\tilde{\phi}\\
	\end{array}
	\right.
	\label{hyps}
\ee
where $\tilde{\phi}$ is the scaled solid fraction given by:

\be
\tilde{\phi}=\frac{\phi_{max}-\phi}{\phi_{max}-\phi_{min}}
\ee
The authors \cite{poulipow05} acknowledged a major difficulty in the application to the vertical chute with no-slip at the walls; shear bands are not finite and of constant width in the quasistatic limit. We have already demonstrated that a Coulomb wall slip criterion can correct this.
In the following we will illustrate the results of our mixing length model including Coulomb wall slip criterion and $F$ and $G$ functions as in Eq. \ref{hyps}, for different chute width and wall roughness.\\
Before that, note the analytical solution achievable in the quasistatic limit, obtained combining Eq. \ref{gapprox} and \ref{hyps}:

\be
\tilde{\phi}=
\left\{
	\begin{array}{ll}
	I_0\frac{y-\mu_s'}{\mu_2'-y}\;\;\;\;\;\;& for\;\;y>\mu_s'\\
	0\;\;\;\;\;\; &for\;\;y\leq \mu_s'\\
	\end{array}
	\right.
\label{phiprof}
\ee
where $\mu'$ are the corresponding parameters in Eq. \ref{hyps} multiplied by $b/tan \delta$. The scaled velocity profile, obtained combining Eq. \ref{shearbandw},\ref{hyps} and \ref{phiprof}, is indeed a simple function of $y$:

\be
\tilde{u}=
\left\{
	\begin{array}{ll}
	A(y-b)+B\;ln(C-Dy) \;\;\;& for\;\;y>\mu_s'\\
	1\;\;\;\;\;\; &for\;\;y\leq \mu_s'\\
	\end{array}
	\right.
\label{velprof}
\ee
where $A$,$B$,$C$,$D$ are known constants, depending on model parameters, wall friction angle and channel width.\\In the following we perform an analysis based on these analytical results; far from the quasistatic limit, one can repeat the calculations using Eq. \ref{gtrue} instead of Eq. \ref{gapprox}, in a numerical fashion.

\subsection{Shear bands thickness and chute width} 
\begin{figure}[!h]
\centering{
\resizebox{0.98\columnwidth}{!}{%
	\includegraphics{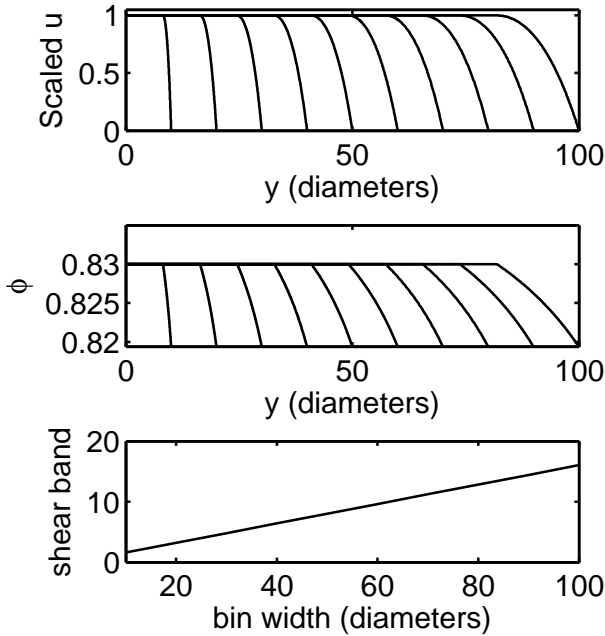}
}}
\caption{Non-dimensional axial velocity, solid fraction and shear bands thickness \emph{vs} non-dimensional channel width, in chutes of different extension. (wall friction angle $\delta=25\deg$).}
	\label{shearboh}
\end{figure}

We already demonstrated that the model predicts shear bands independent by flowrate. Here we explore the effect of bin width.\\
Using the quasistatic assumption, we calculate $\tilde{u}$ from Eq. \ref{velprof}, for different widths, $b$. Other model parameters must be given  and they depend on the specific material choosen. for the purpose of illustrating the model, we used values determined by Jop {\it et al. al.}\cite{jop}, collected in table \ref{parstab}.
\begin{table}[!h]
\caption{Parameters of the model\cite{jop}.}
\label{parstab}
\centering{
\begin{tabular}{lll}
& & \\
\hline\noalign{\smallskip}
$I_0 $ & 0.279 & (adim)\\
\hline
$\mu_s $ & $tan(20.9)$ & (adim)\\
\hline
$\mu_2 $&$ tan(32.76)$& (adim)\\
\noalign{\smallskip}\hline
\end{tabular}}
\end{table}\\
Fig. \ref{shearboh} shows that the model predicts a linear correlation between shear band width and channel width. The slope of the linear dependency may change with different materials, but remains linear. It is frequently stated that the thickness of shear bands is expected to be independent from channel extension. However, literature reports data supporting (e.g. \cite{puligut}) and contrasting \cite{nedderman} this statement. Our results agree with the experimental results by Nedderman and Lahoakul, but the issue requires further investigation of the vertical chute, in order to discriminate the applicability of a mixing-length model to this configuration.

\subsection{Shear bands thickness and wall roughness} 
Wall roughness can influence the extension of the shear bands, according to our model. Adopting Coulomb criterion at the wall yields a simple expression of wall roughness, related to the wall friction angle $\delta$, while using a no-slip condition makes impossible to account for roughness within a continuum approach. Also in view of real scale application of a mixing-length model, wall friction must be correctly accounted for, and the wall friction angle is a wide-spread approach. Furthermore, real world applications aim at perfect wall slip, but are often in an intermediate situation, where wall roughness plays a role.\\

\begin{figure}[!h]
\centering{
\resizebox{0.98\columnwidth}{!}{%
	\includegraphics{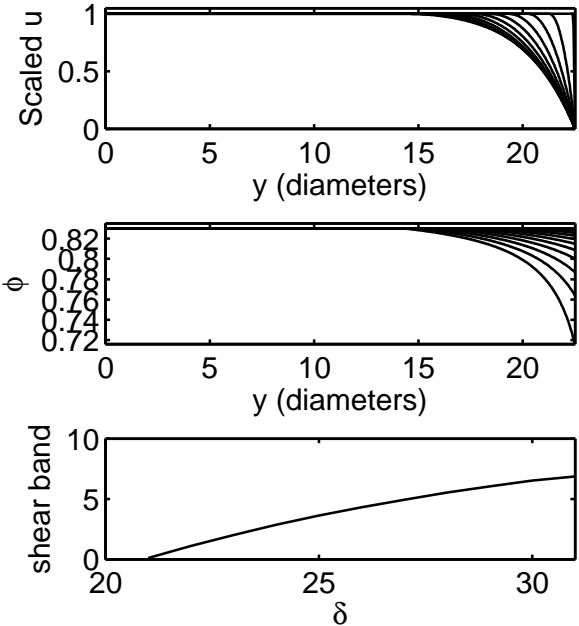}
}}
\caption{
Non-dimensional axial velocity, solid fraction and shear bands thickness \emph{vs} non-dimensional channel width, in chutes with different wall friction (channel width = 50 particle diameters).}
	\label{shearbuh}
\end{figure}

\begin{figure}[!h]
\centering{
\resizebox{0.98\columnwidth}{!}{%
	\includegraphics{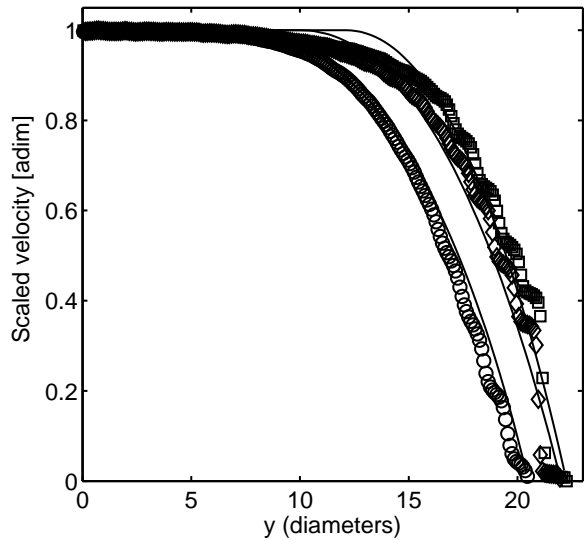}
}}
\caption{Rescaled velocity profiles from DEM simulations by Prochnow \cite{proch} and continuous, mixing-length model. (circles: $D=4\;d$, diamonds: $D=d$, squares: $D=0.5\;d$), while continuous model calculations use different $\delta$ values.}
	\label{shearbih}
\end{figure}

Results according to Eq. \ref{velprof} are given in Fig. \ref{shearbuh}, showing that the model predicts a dependence of shearing regions from wall roughness. However, the effect of $\delta$ is approaching an asymptote. At small values slip occurs, whereas larger friction reduces its influence on the shearing bands. The enlargement shear zones with increasing wall roughness is supported also by DEM results of Prochnow \cite{proch}. Figure \ref{shearbih} illustrates qualitatively the comparison between DEM and our mixing-length, continuous model. The mixing length approach can capture the effect of increasing wall roughness predicted by DEM calculations by means of different sizes of particles fixed at the wall.

\section{Conclusions}
In this work we formulated and applied a mixing-length, compressible continuum model, aiming at discussing shear band particular behaviour in granular flow. 
A general approach based on two functions of the solid fraction to be specified has been given. Application to the vertical chute problem demonstrates that for suitable wall boundary conditions (Coulomb friction), shear band thickness is independent of flowrate, in the quasistatic limit, whatever the form of the generic functions. 
The relevance of these boundary conditions in real applications was discussed, and a critical evaluation of experimental practice was given.\\
Further, we addressed the role of bin width, using expression for the $G(\phi)$ and $F(\phi)$ functions based on reformulation of Literature results \cite{poulipow05} to highlight the variable solid fraction. 
Results do not agree with the frequently assumed independence of shear band thickness from channel width, but agree with with Literature experimental data \cite{nedderman}. We expect that additional experimental investigations might elucidate this controversy.
Taking advantage of the inclusion of wall friction criterion instead of a no-slip boundary condition, we correctly predicted an asymptotic increase of shear bands extension with larger wall roughness. \\
The analysis above proves that even a simplified mixing-length approach to steady granular flows can be useful and representative of a certain number of features, once the proper boundary conditions are used.

\acknowledgments
We acknowledge collaboration by Fran\c{c}ois Chevoir who kindly sent us his DEM results.



\end{document}